# Diffractive X-ray waveguiding reveals orthogonal crystalline stratification in conjugated polymer thin films

*Eliot Gann[1, 2 †], Mario Caironi[3], Yong-Young Noh[4], Yun-Hi Kim[5], and Christopher R McNeill[*1]*

[1] Department of Materials Science and Engineering, Monash University, Wellington Road, Clayton, Victoria, 3800 (Australia)

[2] Australian Synchrotron, 800 Blackburn Road, Clayton, Victoria, 3168 (Australia)

[3] Center for Nano Science and Technology @PoliMi, Istituto Italiano di Tecnologia, Via Pascoli 70/3, Milano 20133 (Italy)

[4] Department of Energy and Materials Engineering, Dongguk University, 26, Pil-dong, 3-ga, Jung-gu, Seoul, 100-715 (Republic of Korea)

[5] Department of Chemistry and RINS, Gyeongsang National University, 501 Jinju Daero, Jinju, 660-701 (Republic of Korea)

The depth dependence of crystalline structure within thin films is critical for many technological applications, but has been impossible to measure directly using common techniques. In this work, by monitoring diffraction peak intensity and location and utilizing the highly angle-dependent waveguiding effects of X-rays near grazing incidence we quantitatively measure the thickness, roughness and orientation of stratified crystalline layers within thin films of a high-performance semiconducting polymer. In particular, this diffractive X-ray waveguiding reveals a self-organized 5-nm-thick crystalline surface layer with crystalline orientation orthogonal to the underlying 65-nm-thick layer. While demonstrated for an organic semiconductor film, this approach is applicable to any thin film material system where stratified crystalline structure and orientation can influence important interfacial processes such as charge injection and field-effect transport.

INTROUCTION

Use of Grazing Incidence Wide Angle X-ray Scattering (GIWAXS) or Grazing Incidence X-ray Diffraction (GIXD/GIXRD) is becoming increasingly popular for characterizing the crystalline structure of thin films, particularly those of organic materials.[1-6] These solution-processed materials often do not produce large crystalline domains, but rather highly paracrystalline nanometer scale domains. These domains are often disordered within the plane of the sample, but have a uniaxial symmetry about the surface, resulting in a uniaxial powder diffraction pattern which is nicely probed by GIWAXS at a single grazing incidence angle. Because these films are macroscopically isotropic in the plane of the sample, the scattering patterns are presented on axes of momentum transfer in the plane of the sample ($Q_{xy}$), and normal to the surface (out of plane $Q_z$). Although characterizing diffraction peaks, because of the high level of paracrystallinity and many crystallites within films, the peaks in GIWAXS of organic thin films are often quite broad, and often the more general term scattering is used to characterize the method rather than diffraction, to distinguish from macro or single crystal diffraction studies, where other techniques are applicable. For further discussion of GIWAXS as applied to organic electronics, the reader is directed to the large body of literature on the subject.[2-3, 7-9] We apply in this work a potentially powerful aspect of GIWAXS that has not been widely utilized, and often ignored in this field: the highly angularly-dependent thin-film X-ray waveguiding effects.[10] As in X-ray reflectivity, fine alteration of the grazing angle of collimated X-rays incident upon a film alters the distribution of X-ray Electric Field Intensity (XEFI) within a film dramatically as shown in **Fig. 1A**, and as utilized in the Distorted Wave Born Approximation (DWBA) treatment of grazing incidence scattering. This framework covers not just the use of evanescent waves, which penetrate into the surface films when below the critical angle, but additionally the interference of incoming and outgoing waves within a thin film above the critical angle. Typical uses of grazing incidence scattering, even analyzed in the DWBA framework use a single or handful or incident angles to look at qualitative surface and bulk features,[3, 11] while experiments such as reflectivity measure many angles but only measure the reflected beam intensity. In a diffractive waveguiding experiment a 2D detector is used to collect scattered X-rays at several different incident angles monitoring diffracted intensities. Because scattering intensity from crystalline structures at a particular depth is directly dependent on the XEFI at that depth, by examining the scattered or diffracted intensity as a function of incident angle the depth-profile of the corresponding morphology can be determined. There are other well-established techniques for depth-profiling of samples, such as Dynamic Secondary Ion Mass Spectroscopy (DSIMS) or X-ray reflectivity, however these techniques are insensitive to the crystalline structure or orientation of different layers. X-ray scattering waveguiding has been demonstrated to reveal non-crystalline nanostructure in thin films,[10-11] and crystalline surface layers of metals.[12] Simple alteration of incident angle to reveal surface structure varying from the bulk of a polymer has a strong history, [13-14] but rarely has a measure of the thickness of surface layers been demonstrated[15] and then only in thick films, where the much simpler penetration depth method is used. In contrast, a full treatment of waveguiding effect in a thin film system has the potential to quantitatively characterize the precise location, thickness, and orientation of the stratified crystalline layers including a precise measure of transition widths between layers.

In cases where angular dependence has been employed to study a surface/bulk difference spin coated conjugated polymer thin films, distinct crystalline surface layers have not been detected, rather only changes in the level of crystallinity have been noted.[16-19] A quantitative measurement of crystalline stratification within thin organic electronic films has yet to be clearly demonstrated.

In self-assembled thin film systems, whether because of surface disorder, film roughness, or lack of highly collimated X-ray beams (with an experimental angular resolution of ~0.0025 degrees), diffractive waveguiding has not been demonstrated. This level of experiment is often judged unnecessary if the primary measurement goal is to examine overall thin film molecular packing, in which case only the angle with the highest overall scattering intensity is analyzed. However in applications such as organic-field effect transistors, where the thin (as little as 2 nm)[20] near-surface layers are responsible for charge transport,[21] simply measuring the average film properties can miss important surface structure that is distinct from the thickness-averaged microstructure.[22]

In this work, we demonstrate unambiguously the crystalline self-reorganization and stratification within a thin film of the novel, high-performance n-type semiconducting polymer poly[(E)-2,7-bis(2-decyltetradecyl)-4-methyl-9-(5-(2-(5-methylselenophen-2-yl)vinyl)selenophen-2-yl)benzo[lmn][3,8] phenanthroline-1,3,6,8(2H,7H)-tetraone] (PNDI-SVS),[23] the chemical structure of which is shown in **Fig. 1B**. PNDI-SVS is representative of high-performance donor-acceptor co-polymers based on the naphthalene diimide (NDI) acceptor unit[24-28] that have demonstrated field-effect electron mobilities of up to 6 cm$^2$/Vs at high voltage[29] making them attractive for application in printed electronics.[29-31] The PNDI-SVS film studied in this investigation was prepared by spin-coating onto a silicon wafer and annealing at 210 °C for 20 minutes, resulting in an electron mobility of 2.2 cm$^2$/Vs in organic field-effect transistors (OFETs) as discussed in detail elsewhere.[23]

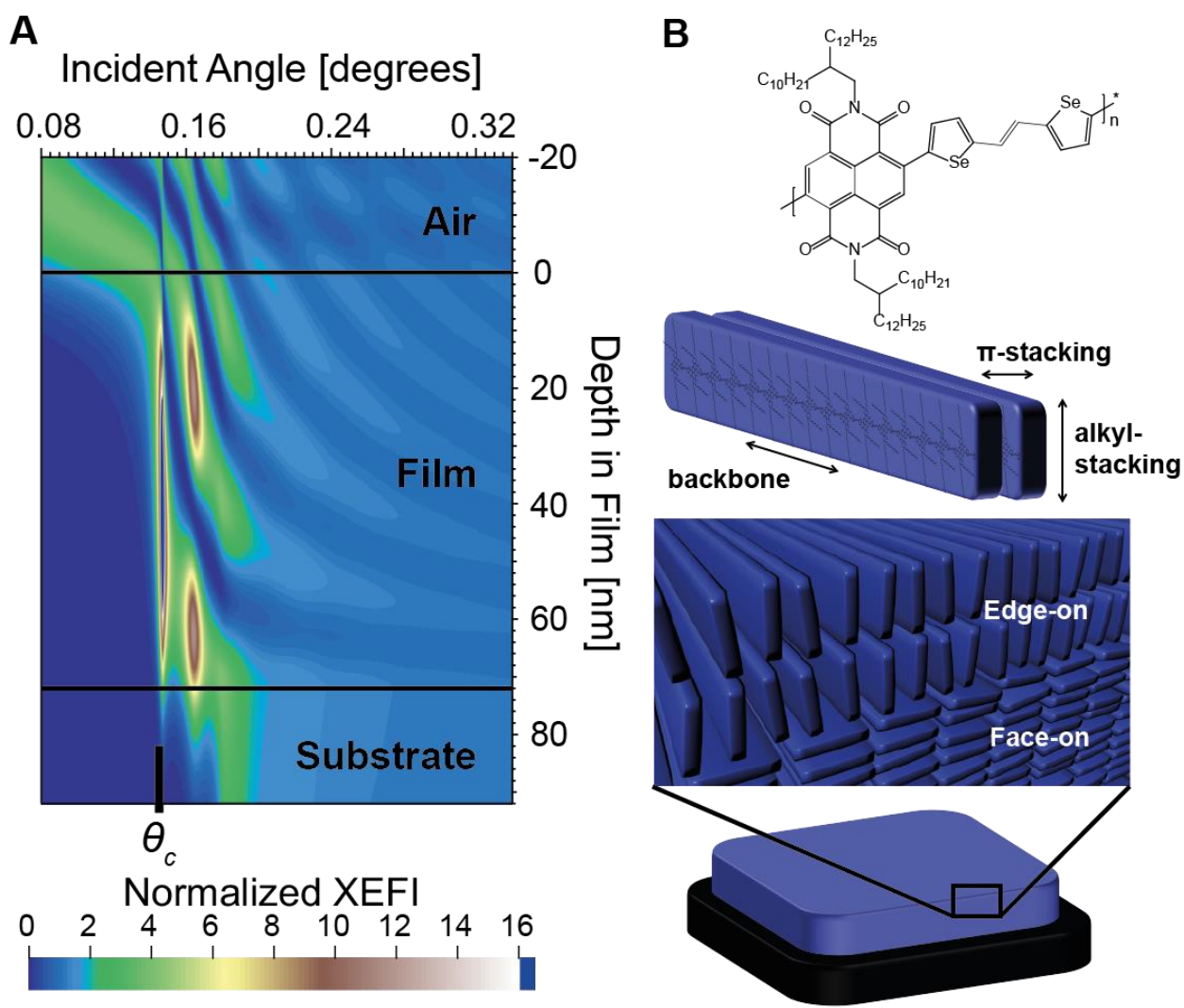

**Figure 1.** Depth Profiling through grazing incidence X-ray scattering and the importance of molecular orientation in organic thin films. A) the XEFI throughout a thin film system consisting of 72 nm of PNDI-SVS on a silicon substrate calculated at 9 keV. B) Chemical structure and schematic of PNDI-SVS molecular packing and stacking in a thin film.

Although semiconducting polymers films are often not highly crystalline because they are produced by solution processing techniques such as spin coating and rely on self-organization into crystalline structures, they do have characteristic paracrystalline packing, as shown in **Fig. 1B**. Packing typically consists of π-stacking from the aromatic face of one polymer to another, alkyl lamella stacking across the alkyl side chains, and the backbone repeat along the rigid polymer chains.[1, 4, 26, 32-35] An edge-on orientation, which allows efficient in-plane transport both across π-stacks as well as along the π-conjugated backbone is the most conducive to high in-plane mobilities,[36] as required for OFET operation. A face on orientation on the other hand would mean layers of alkyl side chains potentially blocking 2D conduction paths for in-plane transport, but would be beneficial for out-of-plane transport important for thin film diodes and solar cells. With several recent reports of face-on-orienting polymers exhibiting high OFET mobilities, there has been recent debate as to whether in real samples, an edge-on orientation is strictly necessary for good OFET performance.[37-38] In some cases, near-edge X-ray absorption fine-structure (NEXAFS) spectroscopy has detected thin edge-on orienting layers on top of an otherwise face-on film,[22] however the thickness and crystallinity of such layers has remained unknown due to the limitations of the NEXAFS technique. The ability to experimentally characterize the crystalline properties of near-surface layers is important for firmly establishing the relationship between surface structure and interfacial processes.

MATERIALS AND METHODS

The synthesis and full device properties of PNDI-SVS are reported elsewhere.[23] The batch of PNDI-SVS used here had a molecular weight (Mw) of 168.0 kg/mol, with a polydispersity index of 2.1.[1] PNDI-SVS films for GIWAXS analysis were prepared by spin-coating onto silicon wafers with a natural oxide layer. PNDI-SVS was dissolved in anhydrous chlorobenzene solvent with a solution concentration of 12 mg/ml. The semiconductor solutions were filtered with a 0.2 µm polytetrafluoroethylene syringe filter and spin coated at 2000 rpm for 60 seconds in N2 filled glove box. The spin coated PNDI-SVS films were thermally annealed at various temperatures for 20 minutes in the N2-purged glove box.

GIWAXS measurements were carried out at the SAXS/WAXS beamline at the Australian Synchrotron[39] with a Dectris Pilatus 1M detector. During different measurements, 9 keV highly collimated photons were aligned parallel to the sample by using a photodiode. Angular steps of 0.01 degrees were taken near the critical angle, which was determined as the angle of maximum scattered intensity. To confirm our results, the measurement was later repeated with 0.005 degree steps (to ensure all angular dependence was captured) and at 11 keV (allowing a larger q range to be obtained) with a similarly prepared sample. 9 keV data is shown in figures 1-2, while the higher resolution 11 keV data is used in the remainder. Each 2D scattering pattern was a result of a total of 3 seconds of exposure. Three 1-second exposures were taken at different detector positions to fill in the gaps between modules in the detector, and combined in software. Correction of data onto momentum transfer axes and sector profiles, was done using an altered version of NIKA.[40] Peak fitting was done using least squares multipeak fitting within IgorPro. The only data correction performed is the change of axes from pixel to in and out of plane q, resulting in the missing wedge due to the incoming and outgoing incident angles differing. For details please refer to extensive GIWAXS/GIXD literature. [1]

XEFI was calculated by Parratt's formalism using a multilayer slicing algorithm[41] within IgorPro. The film is decomposed into a multilayer system, with roughness being accounted for within the system as a gradual change from one index of refraction to another across many Angstrom-thick layers. At each of these layers the incident and reflected beams are reflected and refracted and in each layer absorption is calculated. Setting the boundary conditions of the incoming electric field intensity we solve for the electric field at each interface, allowing determination of the total XEFI, of the incoming and outgoing X-ray wave at each Angstrom of depth within the system. The model ignores in-plane structure, instead averaging the entire system at each depth. Further refinements to the model would take into account the local thickness variations and in-plane structure both of the film as a whole and of the edge-on region, and allow for unaligned regions as well, but these refinements are unnecessary to get a remarkably good agreement between simulation and measurement.

RESULTS

Angle-resolved GIWAXS was performed on PNDI-SVS at the Australian Synchrotron SAXS/WAXS beamline.[42] Here, the very low vertical angular divergence and stable instrumentation allows reliable angular resolution to less than 0.0025 degrees. **Figure 2** shows 2D scattering patterns 0.01 degrees below and above the critical angle ($\theta_c$) of the PNDI-SVS film. The critical angle $\theta_c$ is identified clearly as the angle resulting in a considerable increase in scattering intensity, correspondingly nicely to the angle where we calculate the highest XEFI intensity throughout much of the thin film (as shown in **Fig. 1A** near 0.145°). Figures 2A and B share a color scale, showing the great increase in scattering just above the critical angle for the majority of the peaks. Background structure in the GIWAXS is due to air scatter as the experiments are performed in atmosphere.

The scattering pattern just above $\theta_c$ is similar to that seen in the similarly structured and well-studied P(NDI2OD-T2) polymer[16, 22, 43-44] which substitutes thiophene sub units rather than selenophene units and lacks the vinyl linker in PNDI-SVS. We can identify a largely orthorhombic unit cell as shown schematically in **Fig. 1B**, consisting of strong in-plane (along the $Q_{xy}$ -horizontal axis) alkyl stacking reflections ($h$00) with a spacing of 2.64 ± 0.01 nm and a backbone repeat reflection (00$l$) with a spacing of 1.65 ± 0.01 nm. These in-plane reflections have higher order reflections and coherence lengths of 35 ± 2 nm and 25 ±2 nm for the ($h$00) and (00$l$) reflections respectively. Out of plane, there is a clear π-stacking peak with two monomers per unit cell and a π-π (020) spacing of 0.41 ±0.01 nm and a coherence length of 4 ± 1 nm. All of these major reflections are indicative of face-on polymer orientation within the film. Although difficult to quantify, the crystallinity is relatively high for polymeric semiconductors, with strong well-defined diffraction peaks with several orders and evidence of 3D crystallinity in the mixed index (01$l$) and (023) peaks, which help to allow clear determination of the unit cell.

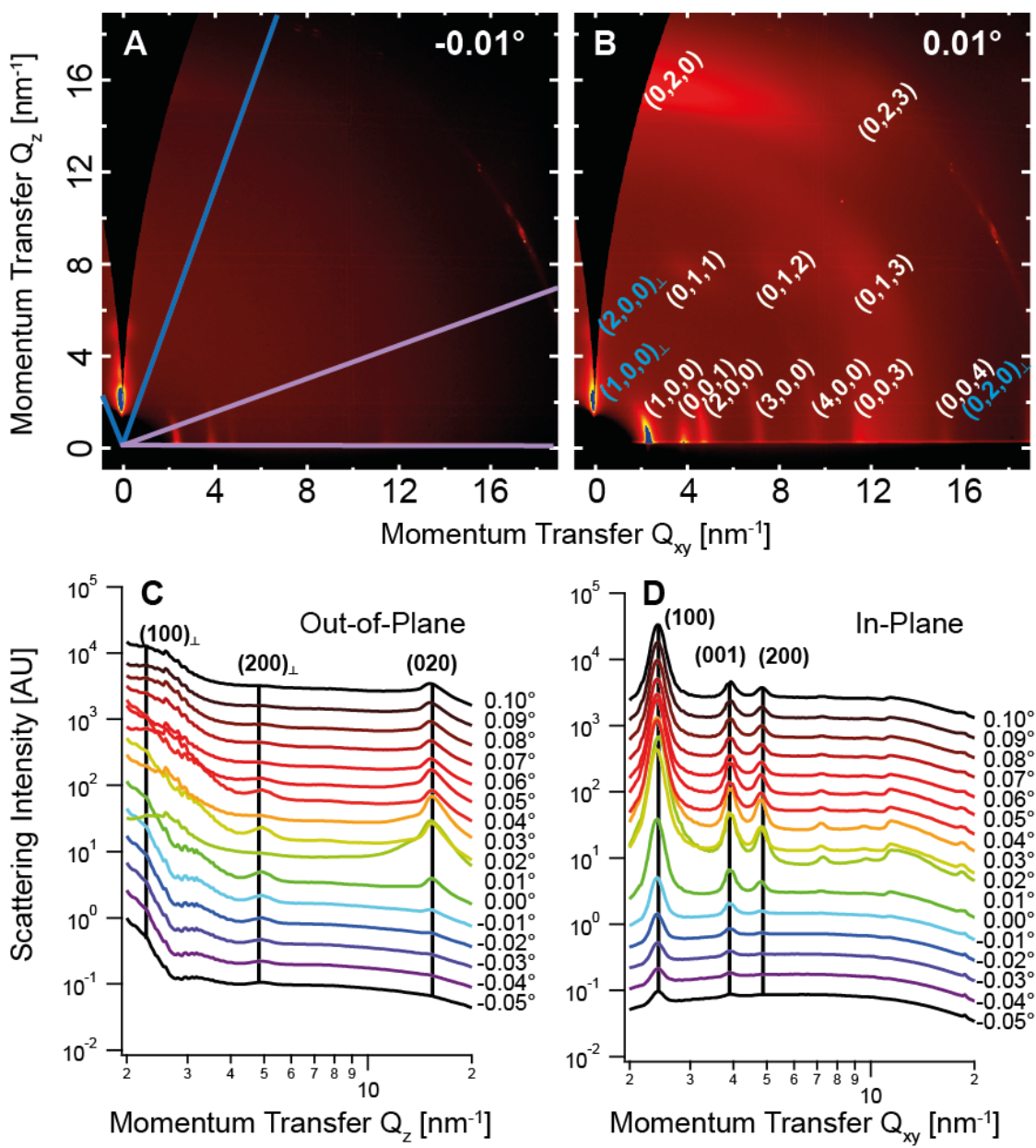

**Figure 2.** 2D GIWAXS patterns of PNDI-SVS, plotted on $Q_z$, $Q_{xy}$ axes, 0.01 degrees below (A) and 0.01 degrees above (B) the critical angle of the polymer. Unit cell indices are shown overlaid over the data in white (face-on oriented lattice) and blue (edge-on oriented lattice). One dimensional profiles for all incident angles from -0.05 to 0.1 degrees (relative to the critical angle) are shown for vertical, out-of-plane (C), and horizontal, in-plane (D) directions. The areas

averaged for the lineouts are shown in (A) as sectors of +/- 20 degrees around the out-of-plane and in-plane directions.

In addition to this face-on crystallinity, there is a second crystalline phase, clearly shown by (h00)⊥ reflections along the $Q_z$ axis (the '⊥' symbol is used to distinguish these crystalline domains that are perpendicular to the more intense reflections from the rest of the film), most obvious is the second order (200)⊥ reflection, which corresponds to a component of the film which appears to be stacking edge-on. The (100)⊥ reflection is difficult to distinguish from the background specular scatter originating from the edges of the substrate and roughness and scratches on the Si wafer. The size of this peak, unlike those corresponding to the edge-on aligned crystallites actually decreases in **Fig. 2B** relative to **Fig. 2A**. The spacing corresponding to the (200)⊥ reflection is 2.55 ± 0.1 nm with a coherence length of 8 ± 3 nm (calculated by Scherrer analysis), in good agreement with the (200) spacing of 2.64 ± 0.01 nm. The higher uncertainty in the out of plane spacing can be understood to be a result of the $Q_z$ uncertainty due to diffracted peaks from the incident and reflected X-ray components[45] in the DWBA, but also a considerably weaker and spread out peak with a smaller coherence length.[11] It is expected that the (00*l*)⊥ reflections exist, but are difficult to reliably pick out of the diffuse background and multiple higher order reflections in that region, while the (0*k*0)⊥ are coincident with the stronger (0*k*0) reflections.

Comparing the angular dependence of the in-plane and out-of-plane scattering we can see that the (100) and (001) face-on reflections compared to the (200)⊥ edge-on reflections (**Fig. 2C and 2D**) have a very different incident-angular dependence, such that the highest intensities occur at different incident angles. This difference indicates that these crystalline orientations must occur at different depths within the thin film. To determine the depth quantitatively, we extract the exact angular dependence of these two crystalline structures. By taking the (100) peak as representative of the face-on component, and fitting the peak as a Gaussian on a log-cubic background, we obtain the peak area in a repeatable fashion. Similarly we take the $(200)_\perp$ reflection, fit in the same manner, as the representation of the edge-on component of the film. We pick these two peaks because they are relatively well-separated from other peaks and backgrounds, such that their peak sizes can be readily fit even when the scattering intensity is low and surrounding and background features are high.

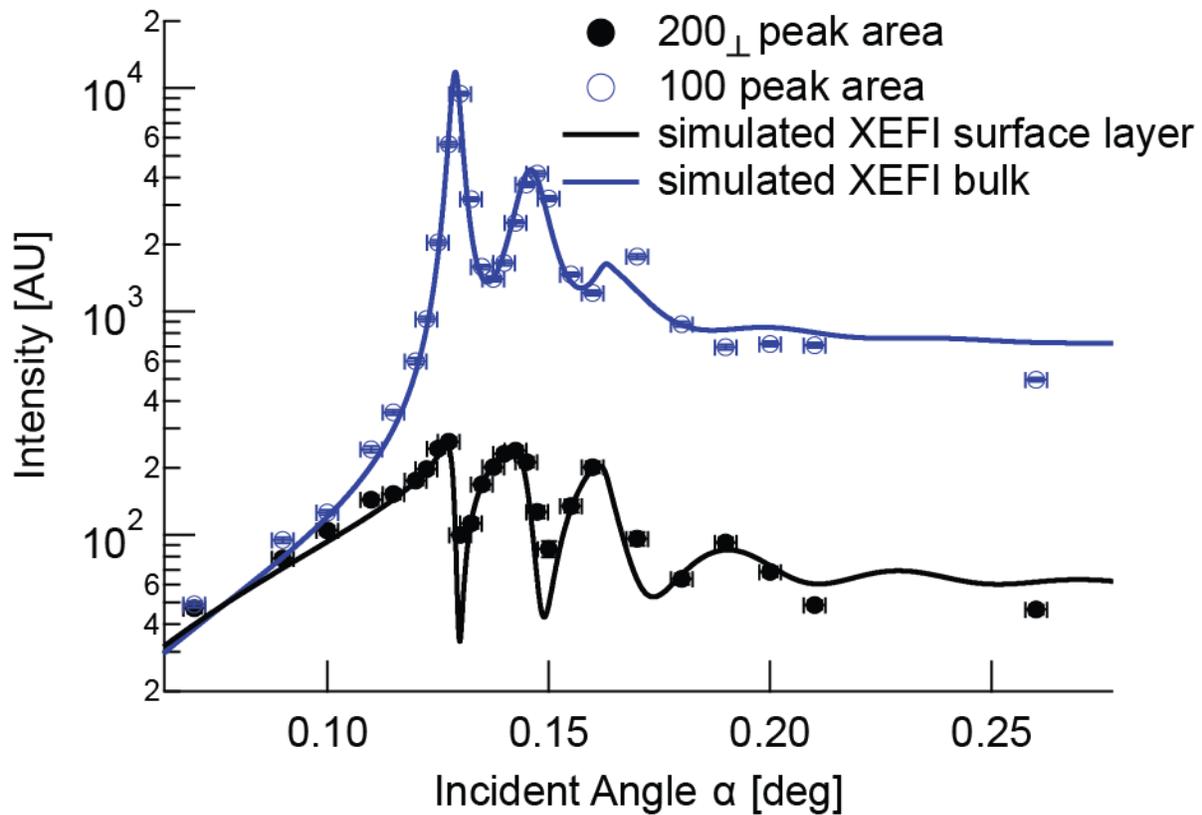

**Figure 3.** The measured angular dependence of the peak areas of the $(200)_\perp$ and (100) reflections, as well as the simulated angular dependence of the XEFI, within a 70 nm film with 2 nm roughness, of the top 5 nm and the bottom 65 nm of the film, with a transition width of 5 nm

between the regions. Uncertainties in intensity come from the peak fit uncertainties, while uncertainty in the incident angle comes from an estimate of the reproducibility of the angular motor stage during acquisition.

These peak-fit areas vs incident angle are plotted in **Fig. 3**. To determine the source within the films which might match these angular profiles, we first setup a simulation of the angular and depth dependent XEFI in the film (as shown previously in **Fig. 1B** and in previous work[10]) for different thicknesses and roughness of films, and then examine the angular profiles from different regions, searching for sets of profiles that match the measured angular dependence. The results of this process is shown in **Fig. 3**, and the effects of slight variations are shown in **Fig. 4**.

Although independent atomic-force microscopy (AFM) measurements of film thickness and roughness were taken (see previous work[23] and **Supporting Information**) we can also narrow down the parameter space using only the diffraction data, as shown in **Fig. 4**. The use of the scattering data to infer surface roughness is particularly important since GIWAXS is sensitive to both surface roughness and film thickness variations (both of which enter into the effective RMS roughness value) over the entire footprint of the beam (~ 20 mm by 0.1 mm), an area which is many orders of magnitude larger than the surface topography measured by AFM. The optimized film parameters plotted in **Fig. 3** determine the film thickness and roughness as well as the different crystalline regions. Particularly, the frequency and relative heights of the maxima and minima force a narrow range of solutions around 70 ± 2 nm thick films, while the initial slopes and ratio of peak heights match the XEFI best for a model with the top 5 nm of the film oriented edge-on and the bottom 65 nm oriented face on with a transition region with a Gaussian width of 5 nm between. Results from simulations for increasing roughness, thicknesses with ± 2 nm, surface regions of depths of ± 2 nm and transition widths of ± 2.5 nm of the above values, shown in **Fig. 4** demonstrate the level of reliability of the optimized results. The extremely close match between measured peak area and simulated XEFI through the critical grazing incidence angular region give confidence that the edge-on crystallites in the PNDI-SVS film are restricted to the top 5 nm of the film and face-on crystallites are largely restricted to the bottom 65 nm of the film. The coherence length of the edge-on component, determined independently by the Scherrer peak width is 8 ± 3 nm, matching the edge-on layer thickness of this model well.

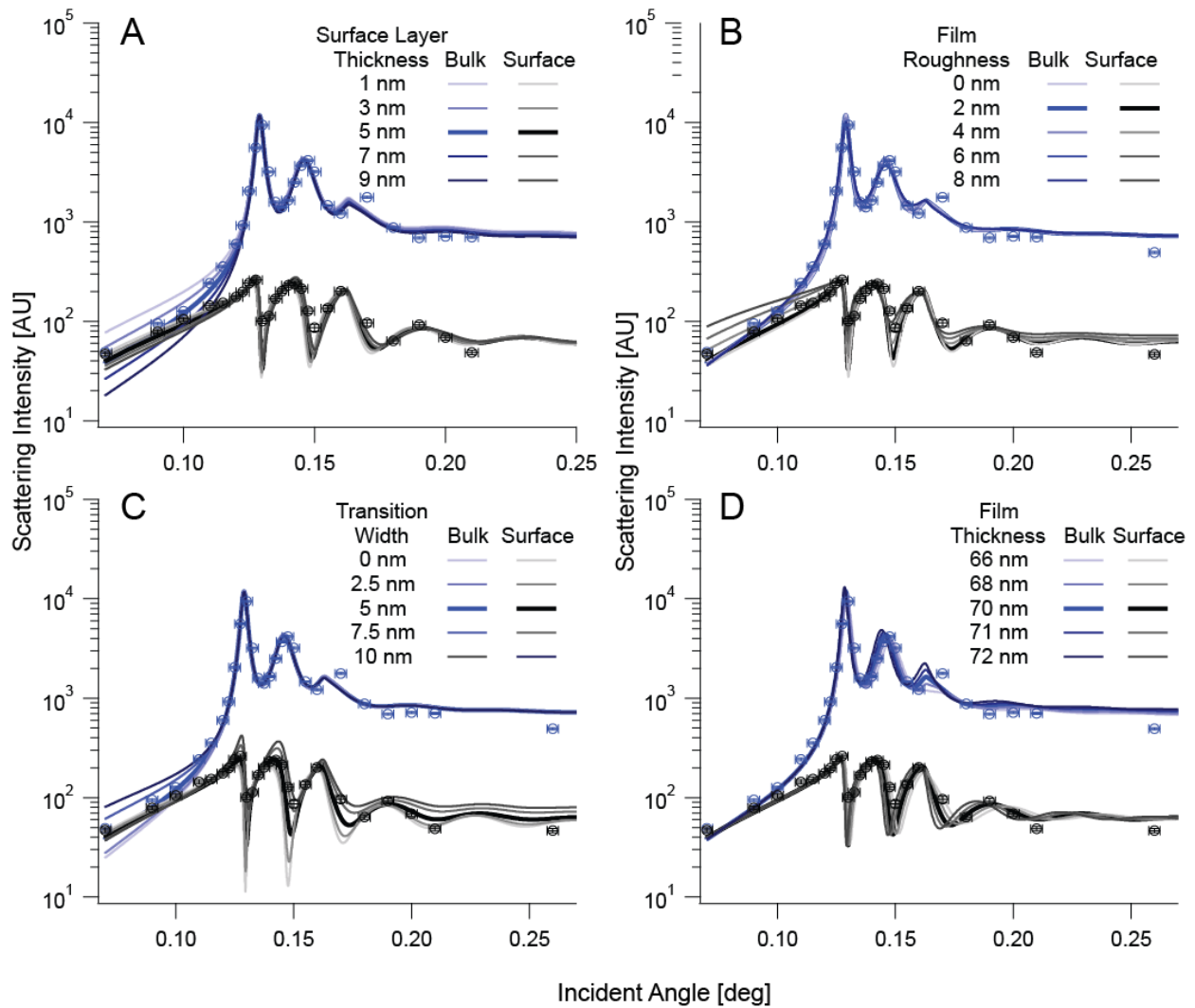

**Figure 4.** Simulated XEFI with slightly differing parameters, compared with the measured intensities. (A) Varying the surface layer thickness from 1 nm, to 9 nm. (B) Varying the film roughness from 0 nm to 8 nm (C) Varying the transition width from 0 to 10 nm, and (D) the total film thickness from 66 nm through 72 nm.

DISCUSSION

These diffractive waveguiding measurements of the PNDI-SVS thin film unambiguously reveal a highly stratified film with distinct crystalline orientations. The thickness and level of roughness/variation of film thickness revealed by diffractive waveguiding match previous AFM results which showed ~1 nm of local surface roughness. The current measurement is sensitive to much longer-range variations, measuring 2 ± 2 nm of thickness variations over the illuminated sample. The AFM additionally revealed a fibrillar morphology and well-defined step heights (previously published and discussed[23]), consistent with a high degree of surface order and high level of edge-on orientation at the surface. This relatively low level of roughness, together with the extremely low divergence of the X-ray beam is important in resolving the details in **Fig. 3**. **Fig. 4,** because higher levels of roughness or higher source divergences would result in a lower angular resolution, which above ~10 nm RMS roughness or 0.01 degrees of divergence would obscure much of the detail necessary for quantitative determination of the film and stratification parameters.

This observation of crystalline orthogonal self-stratification in a solution-processed semiconducting polymer film is a significant observation in its own right, justifying some discussion on the nature and mechanism of this phenomenon. At lower annealing temperatures, an edge-on crystalline component was also found in a majority face-on film, however the angular dependence of the corresponding peaks indicate that that component is spread throughout the film. This can be seen by the lack of significant changes in corresponding peak heights around the critical angle in **Fig. 5** at temperatures lower than 210°C. One possible explaination is that the surface reorganization is at least partially the result of temperature dependent translational mobility of crystallites being slightly easier than either rotational mobility of crystallites or melting of crystals and reforming. Another possibility is that a short anneal at 210 °C just enough thermal energy to melt and reform or rotate isolatedbedge-on crystallites within the film (surrounded on all sides by face-on crystallites) to match their neighbors' orientation, while at the free surface face-on crystallites might have the just enough freedom to begin to reorient to the more energetically favorable edge-on orientation. Electron-yield NEXAFS spectroscopy, which is sensitive to the first few nanometers of the film, confirms a more edge-on (though not necessarily crystalline –NEXAFS measures both crystalline and amorphous material) surface alignment in all spin-coated films even without annealing, while bulk-sensitive fluorescence-yield NEXAFS spectroscopy confirms face-on molecular orientation in the majority of the films.[23] Thus the very surface layer which is consistently edge-on at all temperatures is only able to start to template orientational propagation through the film at 210 °C. And similarly, the bulk of the film which is overwhelmingly face-on at lower temperatures is only able to reorient isolated pockets of edge-on crystals at 210 °C, resulting in a clean stratification.

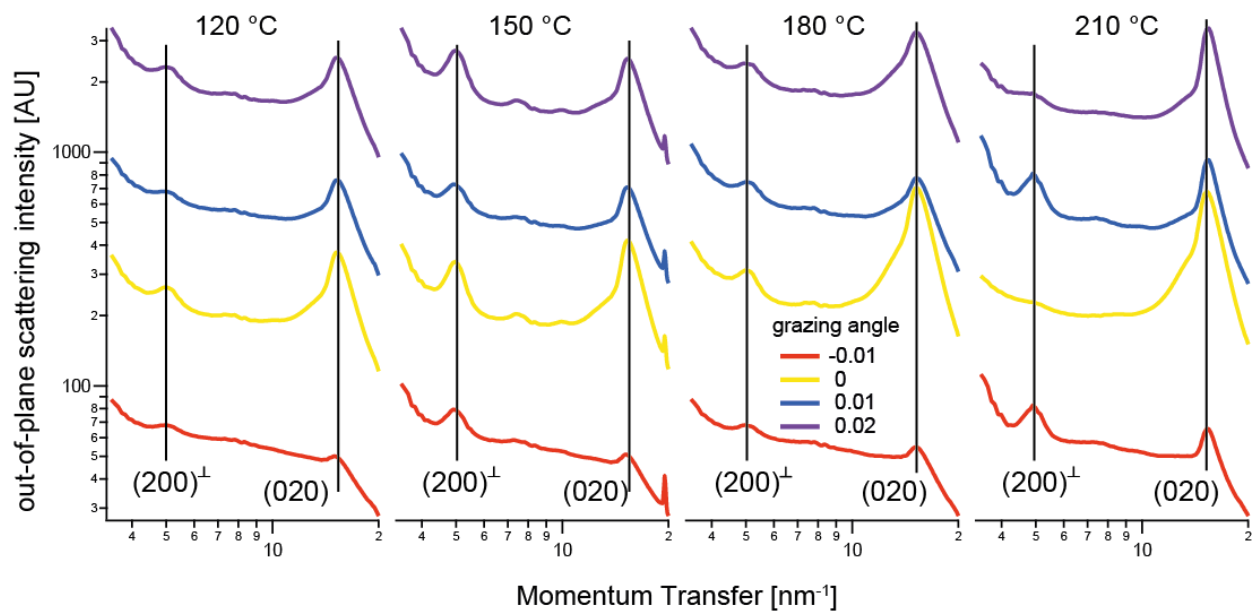

**Figure 5.** Out-of-plane scattering profiles at (red) ~0.01° below the critical angle, (yellow) near the critical angle (blue) ~0.01° above the critical angle and (purple) ~0.02° above the critical angle from films of PNDI-SVS annealed to (left to right) 120 °C, 150 °C, 180 °C, and 210 °C. Note in particular the angular dependence of the $(200)^\perp$ and $(020)$ peaks (pointed out by vertical lines) are basically the same from 120 °C through 180 °C, but changes dramatically at 210 °C, such that the maximum intensity of each peak is at differing grazing angles.

If the sample is annealed for a longer period of time, or at higher temperatures, the edge-on crystallinity does win out in the end, and highly edge-on films are produced,[23] indicating that the stratification is not at equilibrium but perhaps only a snapshot of the process which we kinetically trap upon quenching to room temperature. This observation would lead us to the conclusion that surface energy effects at the free surface are the driving force in the orientation of this organic semiconductor, with the system able to lower its energy by placing the aliphatic side chains at the film/air interface. Indeed other studies have similarly observed a transition from face-on to edge-on with annealing, believed to be templated by edge-on oriented molecules at the film/air interface.[22, 44] Further supporting this observation is a second sample also annealed at 210 °C for a similar amount of time which was found to have a 9 nm surface layer using the same method (see **Supporting Information**), indicating that we may just be catching a relatively slow transition to edge on which is nucleated at the surface and would have proceeded had we not quenched the film. The unique property found in the thin film of PNDI-SVS annealed briefly at 210 °C is that the reorientation is limited to surface layer and that it is crystalline, resulting in the entire film consisting of crystallites in one orientation or the other. We suggest that it is the backbone planarity and close π-π stacking distance and increased backbone stability due to the vinyl linkers between the selenophene groups that allows the high level of film crystallinity and reorientation of backbones on the surface to be minimally hindered by the orientation of the lower levels. This independence allows the surface layer polymers to maintain close enough registration to remain crystalline upon reorienting. Perhaps the morphology of the surface is also important. The fact that the transition layer is so wide implies that some areas the crystalline layer must be thicker, up to 10 nm and in some areas it might be very thin, while the overall film roughness remains much lower.

Relating the observed microstructure to OFET performance, we note that PNDI-SVS films can achieve electron mobilities of ~ 2 cm$^2$/Vs in n-type top-gate OFETs.[23] Interestingly the transport parameters of PNDI-SVS top-gate OFETs did not strongly correlate with "bulk" GIWAXS analysis (that is GIWAXS data that averaged over the entire film thickness), suggesting that molecular orientation and molecular order at the interface determine transport in each sample. Surface sensitive near-edge X-ray absorption fine-structure spectroscopy confirms an edge-on orientation of molecules at the top interface,[23] distinct from the more face-on orientation of the bulk, highlighting the need to use techniques with appropriate surface sensitivity in order to correlate OFET properties with molecular orientation.

Importantly, unlike previous observations of differing surface orientation, the PNDI-SVS sample investigated here shows that the surface layer is crystalline. For other polymers where an edge-on surface has been observed on top of a face-on bulk, a stratified crystalline layer has not been observed. The surface region either fails to crystallize in these other systems, and the reorganization is only seen through NEXAFS, which is sensitive to both amorphous and crystalline molecules,[22] or the surface crystal is too small to have been measured (only a monolayer). As noted above, in many such systems the whole film is seen to change orientation to edge-on upon extended annealing, just as we see in this case.[44] In some cases, it is likely that crystalline stratification was simply not previously noticed, in others it may have been concluded that an edge-on component of scattering was the result of mixed orientation within the film with no further investigation of the subtle angular dependence of the scattering peaks. It is therefore unclear in these previous studies whether an edge-on surface layer is present and non-crystalline, or simply not observed. Surface layers may be too thin to notice or samples may be too rough to allow effective diffractive waveguiding. The fact that high mobilities are also observed in P(NDI2OD-T2)-based top gate transistors suggests that as little as a monolayer of edge-on oriented polymer chains is necessary for good OFET performance, too small to create out of plane diffraction peaks here observed. Even in this case, careful quantification of the background to reveal the angular dependence of in-plane $(0k0)_\perp$ reflections would allow depth profiling through diffractive waveguiding.

Diffractive waveguiding has the potential to characterize films of multiple components, such as organic photovoltaics which have an active layer of a mixture of two organic components. It would be important to consider the refractive index of the different components, which would result in a different EFI calculation for each stratification model in addition to overall film thickness and roughness, which in the present case are not needed because the difference of refractive indices of the stratified layers at hard X-ray energies is negligible.

To understand the potential of diffractive waveguiding to identify more complicated structures than surface segregation and the limitations and benefits of the technique, it is useful to consider the related technique of X-ray reflectivity. Reflectivity is a natural complement to diffractive waveguiding, as in the course of fitting reflectivity profiles, the XEFI distribution throughout the film (as shown in Figure 1A) is solved, which can then be used to determine locations in the film of different crystalline structure. Thus, diffractive waveguiding has the same potential for depth-profiling thin films as X-ray reflectivity, with the distinction that diffractive waveguiding is sensitive to differences in the off-specular crystallographic properties and so can characterize crystalline feature depth, rather than just refractive index depth profiles. It is the diffractive information that reveals the stratification in this work. The same lack of one-to-one correspondence between angular-dependent intensity and depth-profiles that one finds in reflectivity also holds for diffractive waveguiding. As with X-ray reflectivity, the ability to deduce buried structures may require some extra amount of knowledge about the sample for convergence. In the case presented here, an angular resolution of 0.01 degrees and a range of 0.05 to 0.25 degrees would be just sufficient to determine the structure, although the finer resolution of 0.0025 degrees allowed an unambiguous fit. Angular resolution requirements may become finer with higher X-ray energies, and thicker or denser films. A similar requirement is that the thin films need to have low overall roughness to resolve fine features.

It is also important to emphasize that the technique of diffractive waveguiding has great potential for depth-profiling other thin-film systems beyond OFETs and semiconducting polymers. In general, one expects crystalline stratification to produce distinct effects on device physics including charge injection, charge transport, environmental stability, optoelectronic and piezoelectric properties, and refractive properties in myriad materials systems. In any system where different crystalline phases or orientations induce a change in performance, understanding exactly where they lie within a thin film using diffractive waveguiding will be critical to understanding and predicting system performance. We expect that the utilization of X-ray evanescent and waveguiding effects within thin films to unlock a new functionality of grazing X-ray diffraction will be beneficial to many fields where disclosure of depth-dependent crystallinity is important.

CONCLUSIONS

Using angular dependent GIWAXS we have demonstrated the potential for diffractive waveguiding to depth-profile thin film crystallinity. In the first demonstration of this approach, we have revealed the stratification of orthogonal crystalline orientations within a spin coated thin film of a high-performance semiconducting polymer PNDI-SVS, such that the edge-on component is in the top 5 nm of the film, and the face-on component lies in the bottom 65 nm of the film. This observation is made by comparing the angular dependencies of scattering peaks with simulated angular and depth dependent XEFI within a 70-nm thin film. This self-stratification is a novel and potentially important device morphology that can have implications in future organic electronic devices. Additionally, the power of diffractive X-ray waveguiding is demonstrated in a real system, showing the versatility and importance of this tool in characterizing multilayer and complex thin films.

AUTHOR INFORMATION

**Corresponding Author**

**Christopher.McNeill@monash.edu*

**Present Addresses**

† National Institute of Standards and Technology, Gaithersburg, Maryland 20899

**Author Contributions**

The manuscript was written through contributions of all authors. All authors have given approval to the final version of the manuscript.

**Funding Sources**

C.R.M. acknowledges funding from the Australian Research Council (DP130102616). M.C. acknowledges final support from the European Research Council (ERC) under the European Union's Horizon 2020 research and innovation programme "HEROIC", grant agreement 638059.

ACKNOWLEDGMENT

This work was undertaken in part at the SAXS/WAXS beamline of the Australian Synchrotron, Victoria, Australia. E.G. and C.R.M. thank Stephen Mudie for assistance setting up the GIWAXS measurements. The authors acknowledge Alessandro Luzio for the preparation of the thin films.

ABBREVIATIONS

OFET, Organic Field Effect Transistor; NEXAFS, Near Edge X-ray Absorption Fine Structure; GIWAXS, Grazing Incidence Wide Angle X-ray Scattering; XEFI / EFI, X-ray Electric Field Intensity; AFM, Atomic Force Microscope; RMS, root mean squared; DWBA, Distorted Wave Born Approximation; GIXD/GIXRD, Grazing Incidence X-ray Diffraction

REFERENCES

1. Hexemer, A.; Müller-Buschbaum, P., Advanced grazing-incidence techniques for modern soft-matter materials analysis. *IUCrJ* **2015,** *2* (Pt 1), 106-125.
2. Kline, R. J.; McGehee, M. D.; Toney, M. F., Highly oriented crystals at the buried interface in polythiophene thin-film transistors. *Nat. Mater.* **2006,** *5* (3), 222-228.
3. Müller-Buschbaum, P., The Active Layer Morphology of Organic Solar Cells Probed with Grazing Incidence Scattering Techniques. *Adv. Mater.* **2014,** *26* (46), 7692-709.
4. Lu, X.; Hlaing, H.; Germack, D. S.; Peet, J.; Jo, W. H.; Andrienko, D.; Kremer, K.; Ocko, B. M., Bilayer order in a polycarbazole-conjugated polymer. *Nat Commun* **2012,** *3*, 795.
5. McCulloch, I.; Heeney, M.; Bailey, C.; Genevicius, K.; MacDonald, I.; Shkunov, M.; Sparrowe, D.; Tierney, S.; Wagner, R.; Zhang, W.; Chabinyc, M. L.; Kline, R. J.; McGehee, M. D.; Toney, M. F., Liquid-crystalline semiconducting polymers with high charge-carrier mobility. *Nat. Mater.* **2006,** *5* (4), 328-333.
6. Schubert, M.; Collins, B. A.; Mangold, H.; Howard, I. A.; Schindler, W.; Vandewal, K.; Roland, S.; Behrends, J.; Kraffert, F.; Steyrleuthner, R.; Chen, Z.; Fostiropoulos, K.; Bittl, R.; Salleo, A.; Facchetti, A.; Laquai, F.; Ade, H. W.; Neher, D., Correlated Donor/Acceptor Crystal Orientation Controls Photocurrent Generation in All-Polymer Solar Cells. *Adv. Funct. Mater.* **2014,** *24* (26), 4068-4081.
7. Kim, Y.; Cook, S.; Tuladhar, S. M.; Choulis, S. A.; Nelson, J.; Durrant, J. R.; Bradley, D. D. C.; Giles, M.; McCulloch, I.; Ha, C.-S.; Ree, M., A strong regioregularity effect in self-organizing conjugated polymer films and high-efficiency polythiophene:fullerene solar cells. *Nat. Mater.* **2006,** *5* (3), 197-203.
8. DeLongchamp, D. M.; Kline, R. J.; Jung, Y.; Germack, D. S.; Lin, E. K.; Moad, A. J.; Richter, L. J.; Toney, M. F.; Heeney, M.; McCulloch, I., Controlling the Orientation of Terraced Nanoscale "Ribbons" of a Poly(thiophene) Semiconductor. *ACS Nano* **2009,** *3* (4), 780-787.

9. Baker, J. L.; Jimison, L. H.; Mannsfeld, S.; Volkman, S.; Yin, S.; Subramanian, V.; Salleo, A.; Alivisatos, A. P.; Toney, M. F., Quantification of Thin Film Crystallographic Orientation Using X-ray Diffraction with an Area Detector. *Langmuir* **2010,** *26* (11), 9146-9151.
10. Jiang, Z.; Lee, D. R.; Narayanan, S.; Wang, J.; Sinha, S. K., Waveguide-enhanced grazing-incidence small-angle x-ray scattering of buried nanostructures in thin films. *Phys. Rev. B* **2011,** *84* (7), 075440.
11. Renaud, G.; Lazzari, R.; Leroy, F., Probing surface and interface morphology with Grazing Incidence Small Angle X-Ray Scattering. *Surf. Sci. Rep.* **2009,** *64* (8), 255-380.
12. Robinson, I. K., Direct Determination of the Au(110) Reconstructed Surface by X-Ray Diffraction. *Phys. Rev. Lett.* **1983,** *50* (15), 1145-1148.
13. Yakabe, H.; Sasaki, S.; Sakata, O.; Takahara, A.; Kajiyama, T., Paracrystalline Lattice Distortion in the Near-Surface Region of Melt-Crystallized Polyethylene Films Evaluated by Synchrotron-Sourced Grazing-Incidence X-ray Diffraction. *Macromolecules* **2003,** *36* (16), 5905-5907.
14. Yakabe, H.; Tanaka, K.; Nagamura, T.; Sasaki, S.; Sakata, O.; Takahara, A.; Kajiyama, T., Grazing Incidence X-ray Diffraction Study on Surface Crystal Structure of Polyethylene Thin Films. *Polym. Bull.* **2005,** *53* (3), 213-222.
15. Factor, B. J.; Russell, T. P.; Toney, M. F., Surface-induced ordering of an aromatic polyimide. *Phys. Rev. Lett.* **1991,** *66* (9), 1181-1184.
16. Schuettfort, T.; Huettner, S.; Lilliu, S.; Macdonald, J. E.; Thomsen, L.; McNeill, C. R., Surface and Bulk Structural Characterization of a High-Mobility Electron-Transporting Polymer. *Macromolecules* **2011,** *44* (6), 1530-1539.
17. Tong, M.; Cho, S.; Rogers, J. T.; Schmidt, K.; Hsu, B. B. Y.; Moses, D.; Coffin, R. C.; Kramer, E. J.; Bazan, G. C.; Heeger, A. J., Higher Molecular Weight Leads to Improved Photoresponsivity, Charge Transport and Interfacial Ordering in a Narrow Bandgap Semiconducting Polymer. *Adv. Funct. Mater.* **2010,** *20* (22), 3959-3965.
18. Wang, T.; Pearson, A.; Dunbar, A. F.; Staniec, P.; Watters, D.; Coles, D.; Yi, H.; Iraqi, A.; Lidzey, D.; Jones, R. L., Competition between substrate-mediated π-π stacking and surface-mediated Tg depression in ultrathin conjugated polymer films. *The European Physical Journal E* **2012,** *35* (12), 9807.
19. Porzio, W.; Scavia, G.; Barba, L.; Arrighetti, G.; McNeill, C. R., On the packing and the orientation of P(NDI2OD-T2) at low molecular weight. *Eur. Polym. J.* **2014,** *61* (0), 172-185.
20. Maddalena, F.; de Falco, C.; Caironi, M.; Natali, D., Assessing the width of Gaussian density of states in organic semiconductors. *Org. Electron.* **2015,** *17*, 304-318.
21. Sirringhaus, H., Device Physics of Solution-Processed Organic Field-Effect Transistors. *Adv. Mater.* **2005,** *17* (20), 2411-2425.
22. Schuettfort, T.; Thomsen, L.; McNeill, C. R., Observation of a Distinct Surface Molecular Orientation in Films of a High Mobility Conjugated Polymer. *J. Am. Chem. Soc.* **2012,** *135* (3), 1092-1101.
23. Sung, M. J.; Luzio, A.; Park, W. T.; Kim, R.; Gann, E.; Maddalena, F.; Pace, G.; Xu, Y.; Natali, D.; de Falco, C.; Dang, L.; McNeill, C. R.; Caironi, M.; Noh, Y. Y.; Kim, Y. H., High-Mobility Naphthalene Diimide and Selenophene-Vinylene-Selenophene-Based Conjugated Polymer: n-Channel Organic Field-Effect Transistors and Structure-Property Relationship. *Adv. Funct. Mater.* **2016,** *26* (27), 4984-4997.
24. Hwang, Y. J.; Courtright, B. A.; Ferreira, A. S.; Tolbert, S. H.; Jenekhe, S. A., 7.7% Efficient All-Polymer Solar Cells. *Adv. Mater.* **2015,** *27* (31), 4578-84.

25. Hu, Y.; Gao, X.; Di, C.-a.; Yang, X.; Zhang, F.; Liu, Y.; Li, H.; Zhu, D., Core-Expanded Naphthalene Diimides Fused with Sulfur Heterocycles and End-Capped with Electron-Withdrawing Groups for Air-Stable Solution-Processed n-Channel Organic Thin Film Transistors. *Chem. Mater.* **2011,** *23* (5), 1204-1215.
26. Zhang, F.; Hu, Y.; Schuettfort, T.; Di, C.-a.; Gao, X.; McNeill, C. R.; Thomsen, L.; Mannsfeld, S. C. B.; Yuan, W.; Sirringhaus, H.; Zhu, D., Critical Role of Alkyl Chain Branching of Organic Semiconductors in Enabling Solution-Processed N-Channel Organic Thin-Film Transistors with Mobility of up to 3.50 cm2 V–1 s–1. *J. Am. Chem. Soc.* **2013,** *135* (6), 2338-2349.
27. Szumilo, M. M.; Gann, E. H.; McNeill, C. R.; Lemaur, V.; Oliver, Y.; Thomsen, L.; Vaynzof, Y.; Sommer, M.; Sirringhaus, H., Structure Influence on Charge Transport in Naphthalenediimide–Thiophene Copolymers. *Chem. Mater.* **2014,** *26* (23), 6796-6804.
28. Yan, H.; Chen, Z. H.; Zheng, Y.; Newman, C.; Quinn, J. R.; Dotz, F.; Kastler, M.; Facchetti, A., A high-mobility electron-transporting polymer for printed transistors. *Nature* **2009,** *457* (7230), 679-686.
29. Bucella, S. G.; Luzio, A.; Gann, E.; Thomsen, L.; McNeill, C. R.; Pace, G.; Perinot, A.; Chen, Z.; Facchetti, A.; Caironi, M., Macroscopic and high-throughput printing of aligned nanostructured polymer semiconductors for MHz large-area electronics. *Nat Commun* **2015,** *6*, 8394.
30. Coropceanu, V.; Cornil, J.; da Silva Filho, D. A.; Olivier, Y.; Silbey, R.; Bredas, J. L., Charge transport in organic semiconductors. *Chem. Rev.* **2007,** *107* (4), 926-952.
31. Anthony, J. E.; Facchetti, A.; Heeney, M.; Marder, S. R.; Zhan, X., n-Type organic semiconductors in organic electronics. *Adv. Mater.* **2010,** *22* (34), 3876-92.
32. Hammond, M. R.; Kline, R. J.; Herzing, A. A.; Richter, L. J.; Germack, D. S.; Ro, H. W.; Soles, C. L.; Fischer, D. A.; Xu, T.; Yu, L. P.; Toney, M. F.; DeLongchamp, D. M., Molecular Order in High-Efficiency Polymer/Fullerene Bulk Heterojunction Solar Cells. *Acs Nano* **2011,** *5* (10), 8248-8257.
33. O'Connor, B. T.; Reid, O. G.; Zhang, X.; Kline, R. J.; Richter, L. J.; Gundlach, D. J.; DeLongchamp, D. M.; Toney, M. F.; Kopidakis, N.; Rumbles, G., Morphological Origin of Charge Transport Anisotropy in Aligned Polythiophene Thin Films. *Adv. Funct. Mater.* **2014,** *24* (22), 3422-3431.
34. Gann, E.; Gao, X.; Di, C. a.; McNeill, C. R., Phase Transitions and Anisotropic Thermal Expansion in High Mobility Core-expanded Naphthalene Diimide Thin Film Transistors. *Adv. Funct. Mater.* **2014,** *24* (45), 7211-7220.
35. Rivnay, J.; Noriega, R.; Kline, R. J.; Salleo, A.; Toney, M. F., Quantitative analysis of lattice disorder and crystallite size in organic semiconductor thin films. *Phys. Rev. B* **2011,** *84* (4), 045203.
36. O'Connor, B.; Kline, R. J.; Conrad, B. R.; Richter, L. J.; Gundlach, D.; Toney, M. F.; DeLongchamp, D. M., Anisotropic Structure and Charge Transport in Highly Strain-Aligned Regioregular Poly(3-hexylthiophene). *Adv. Funct. Mater.* **2011,** *21* (19), 3697-3705.
37. Gargi, D.; Kline, R. J.; DeLongchamp, D. M.; Fischer, D. A.; Toney, M. F.; O'Connor, B. T., Charge Transport in Highly Face-On Poly(3-hexylthiophene) Films. *The Journal of Physical Chemistry C* **2013,** *117* (34), 17421-17428.
38. Fabiano, S.; Musumeci, C.; Chen, Z.; Scandurra, A.; Wang, H.; Loo, Y.-L.; Facchetti, A.; Pignataro, B., From Monolayer to Multilayer N-Channel Polymeric Field-Effect Transistors with Precise Conformational Order. *Adv. Mater.* **2012,** *24* (7), 951-956.

39. Cowie, B. C. C.; Tadich, A.; Thomsen, L., The Current Performance of the Wide Range (90--2500 eV) Soft X-ray Beamline at the Australian Synchrotron. *AIP Conf. Proc.* **2010,** *1234* (1), 307-310.
40. Ilavsky, J., Nika: software for two-dimensional data reduction. *J. Appl. Crystallogr.* **2012,** *45* (2), 324-328.
41. Parratt, L. G., Surface Studies of Solids by Total Reflection of X-rays. *Phys. Rev.* **1954,** *95* (2), 359.
42. Kirby, N. M.; Mudie, S. T.; Hawley, A. M.; Cookson, D. J.; Mertens, H. D. T.; Cowieson, N.; Samardzic-Boban, V., A low-background-intensity focusing small-angle X-ray scattering undulator beamline. *J. Appl. Crystallogr.* **2013,** *46* (6), 1670-1680.
43. Rivnay, J.; Toney, M. F.; Zheng, Y.; Kauvar, I. V.; Chen, Z.; Wagner, V.; Facchetti, A.; Salleo, A., Unconventional Face-On Texture and Exceptional In-Plane Order of a High Mobility n-Type Polymer. *Adv. Mater.* **2010,** *22* (39), 4359-4363.
44. Rivnay, J.; Steyrleuthner, R.; Jimison, L. H.; Casadei, A.; Chen, Z. H.; Toney, M. F.; Facchetti, A.; Neher, D.; Salleo, A., Drastic Control of Texture in a High Performance n-Type Polymeric Semiconductor and Implications for Charge Transport. *Macromolecules* **2011,** *44* (13), 5246-5255.
45. Resel, R.; Bainschab, M.; Pichler, A.; Dingemans, T.; Simbrunner, C.; Stangl, J.; Salzmann, I., Multiple scattering in grazing-incidence X-ray diffraction: impact on lattice-constant determination in thin films. *Journal of Synchrotron Radiation* **2016,** *23* (3), 729-734.

BRIEFS (Word Style "BH_Briefs"). If you are submitting your paper to a journal that requires a brief, provide a one-sentence synopsis for inclusion in the Table of Contents.

SYNOPSIS (Word Style "SN_Synopsis_TOC"). If you are submitting your paper to a journal that requires a synopsis, see the journal's Instructions for Authors for details.